\documentclass[10pt,letterpaper,twocolumn]{article} %% two column, final layout

\usepackage{ol2}
\usepackage[draft]{hyperref}
\usepackage{amsmath}

\begin{document}

\twocolumn[ %% activate for two-column option

\title{Photonic Flatband Lasers}

%% For REVTeX it is possible to automate superscript and e-mail callouts with the superscriptaddress option; see REVTeX4 documentation.

\author{Stefano Longhi}

\address{Dipartimento di Fisica, Politecnico di Milano and Istituto di Fotonica e Nanotecnologie del Consiglio Nazionale delle Ricerche, Piazza L. da Vinci 32, I-20133 Milano, Italy (stefano.longhi@polimi.it)}

\begin{abstract}
 Flatband photonic lattices, i.e. arrays of waveguides or resonators displaying a flat Bloch band, offer new routes for light trapping and distortion-free imaging. Here it is shown that flatland lattices can show stable and cooperative laser emission when optical gain is supplied to the system, despite the large degree of degeneracy of flatland supermodes. By considering a quasi one-dimensional rhombic lattice of coupled semiconductor microrings, selective pumping of the outer sublattices can induce cooperative lasing in a supermode of the flat band. 
\end{abstract}

%\ocis{130.3120, 290.5839, 000.1600}
 ] %% activate for two-column option

{\it Introduction}. Flatband photonic lattices, i.e. arrays of coupled waveguides or resonators displaying at least one completely flat Bloch band, have attracted a great interest in recent years \cite{r1,r2,r3,r4,r5,r6,r7,r8,r9,r10,r11,r12,r13,r14,r15}. Inspired by certain lattice models earlier introduced in solid-state physics \cite{r16,r17}, photonic flatland structures  find analogies  with  $^{\prime}$frustrated$^{\prime}$ condensed matter systems and can find potential applications to diffractionless propagation, light trapping, imaging and slow light (see \cite{r1,r2} for recent reviews). Most of photonic flat band systems considered so far are passive structures, i.e. without optical gain, however interesting effects have been recently predicted  to arise in non-Hermitian flat band systems involving optical gain and loss \cite{r18,r19,r20,r21}. In a different yet related context, proposals and experimental demonstrations of topological lasers, i.e.lasers where the cavity is a photonic topological insulator sustaining edge states, have raised a lively attention recently \cite{r22,r23,r24,r25,r26,r27,r28,r29,r30,r30bis,r30tris,r31}. Arrays of coupled semiconductor microring lasers provide a promising platform for the realization of integrated topological lasers \cite{r26,r27,r28}. A natural question than arises: can a photonic flat band lattice of coupled resonators operates as a laser when optical gain is supplied to the system? At first sight, owing to the strong degeneracy of flat bands enabling the formation of clusters of localized modes in the structure, cooperative lasing from the entire system is expected to be prevented, especially when detrimental dynamical instabilities are considered \cite{r30,r30bis,r30tris,r32,r33,r34}. Contrary to such a wisdom, in this Letter it is shown that stable and synchronized laser emission can be attained in flat band lattices with optical gain. We consider a quasi one-dimensional (1D) rhombic lattice of semiconductor microring resonators, sustaining a flat band \cite{r3,r6,r9,r14,r15}, and show that, under selective pumping of the flat band modes, laser oscillation in a collective supermode of the flat band, corresponding to synchronized laser emission, can be observed.\par
 \begin{figure}[htb]
\centerline{\includegraphics[width=8.4cm]{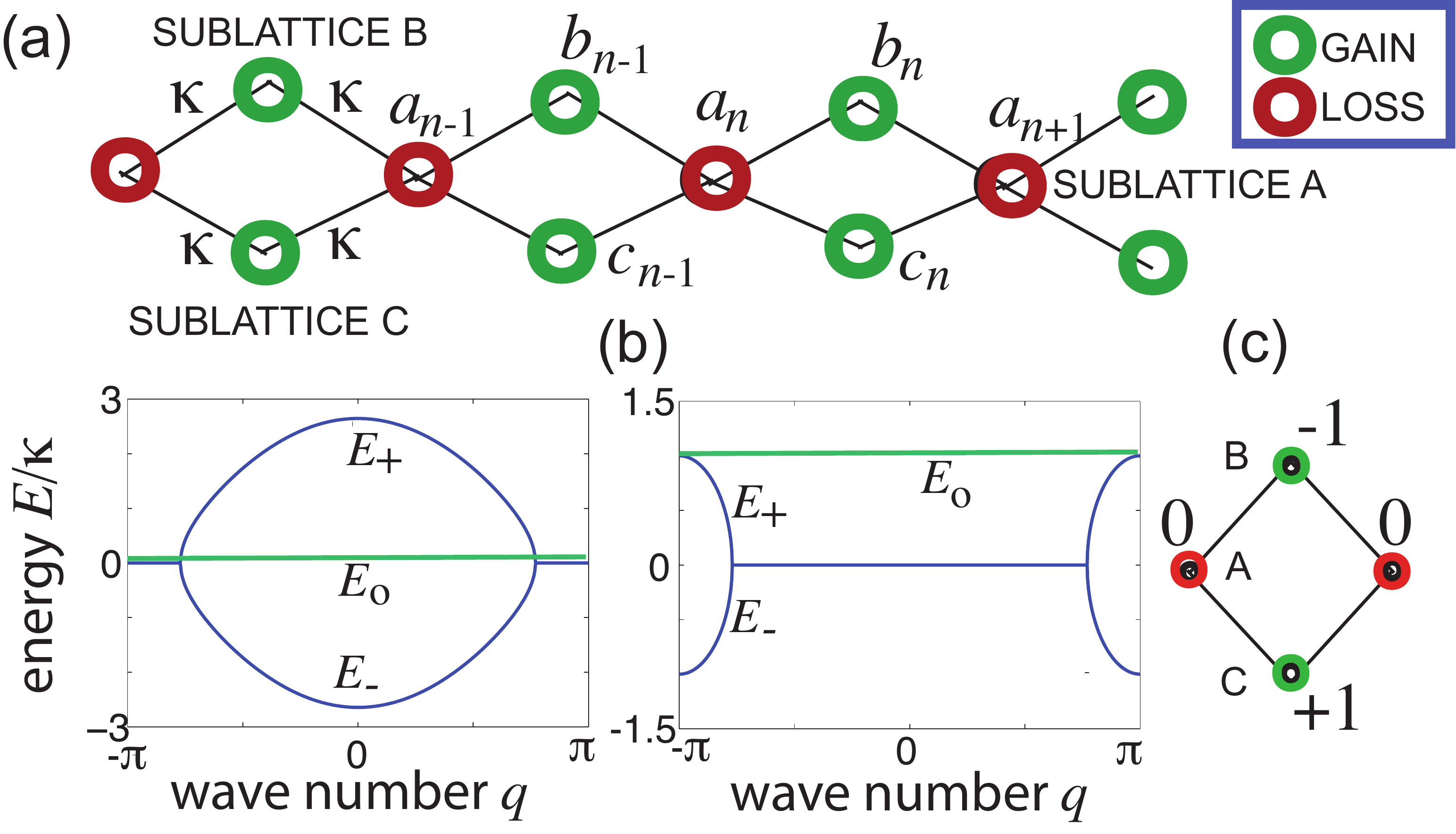}} \caption{ \small
(Color online) (a) Schematic of a rhombic lattice of coupled microrings, comprising three sublattices A,B and C, sustaining a flat band. Optical gain is applied to the outer rings (sublattices B and C), whereas the rings in the sublattice A are lossy. (b) Band dispersion curves (left plot: real part; right plot: imaginary part) for $g=\gamma= \kappa$. The flat band corresponds to the existence of compact (dark) states which localize excitation in two microrings of sublattice B and C [panel (c)], which decouple from the lattice because of destructive interference.}
\end{figure} 
 {\it Flat-band laser: supermode analysis}.
  We consider a quasi 1D  rhombic lattice of evanescently-coupled semiconductor microrings, schematically shown in Fig.1(a), which is known to sustain one flat band in the Hermitian limit \cite{r3,r6,r9}. To realize a flatband laser, we consider optical gain $g$ in the outer sublattices B and C, and optical loss $\gamma$ in the inner sublattice A [Fig.1(a)]. Note that the condition $g=\gamma$ corresponds to parity-time ($\mathcal{PT}$) symmetric excitation \cite{r27}. This pumping scheme favors oscillation of modes in sublattices B and C, i.e. of flatland supermodes. Assuming the same resonance frequency of microrings and single-longitudinal unidirectional mode operation, coupled mode equations describing light dynamics in the lattice read \cite{r26,r27}
  \begin{eqnarray}
  i \frac{da_n}{dt} & = & \kappa (b_{n-1}+b_n+c_{n-1}+c_n)-i \gamma a_n \\
  i \frac{db_n}{dt} & = & \kappa (a_{n+1}+a_n)+i g b_n \; , \;\; i \frac{dc_n}{dt}  =  \kappa (a_{n+1}+a_n)+i g c_n \nonumber
  \end{eqnarray}  
  where $\kappa$ is the coupling constant between adjacent rings and $a_n$, $b_n$, $c_n$ are the amplitudes of modal fields circulating in the $n$-th microring of sublattices A,B and C, respectively. In the linear regime, where $g$ and $\gamma$ are constant parameters, the lattice bands are obtained by making the Ansatz $(a_n,b_n,c_n)^T=(A,B,C)^T \exp(iqn-iEt)$ is Eq.(1), where $q$ is the Bloch wave number. This yields the following dispersion curves of the three lattice bands
  \begin{equation}
  E_{\pm}(q)=i \frac{g-\gamma}{2} \pm 2 \kappa \sqrt{1+\cos q- \left( \frac{g+\gamma}{4 \kappa}\right)^2} \;, \; E_0(q)=i g
  \end{equation}
  A typical behavior of the bands is shown in Fig.1(b) in the $\mathcal{PT}$ symmetric case $g=\gamma$. The $E_0(q)$ curve corresponds to the flat band, with the largest gain over the other dispersive two bands $E_{\pm}(q)$.  The flat band  is related to the existence of a set of complex-energy $E_0=ig$ compact (dark) states, with localization in two sites of the outer sublattices B and C which decouple from the chain owing to  destructive interference [Fig.1(c)]. In the following, we assume that the lattice comprises a number $N$ of unit cells and assume periodic boundary conditions $a_{n+N},b_{n+N},c_{n+N}=a_n,b_n,c_n$, corresponding to a closed lattice on a ring \cite{r33,r34}. The periodic boundary conditions yield a quantization of the Bloch wave number $q=q_l= 2 \pi l/N$ ($l=0,1,2,...,N-1$), 
 where $l$ is the supermode index. Note that all supermodes belonging to the flat band experience the same gain $g$, and thus are degenerate in threshold. This is in contrast to topological laser structures, where a single topologically-protected edge mode reaches threshold \cite{r26,r27,r28}. \par 
  
 {\it Laser rate equations analysis.}  Owing to the high degree of threshold degeneracy of supermodes and the existence of compact dark states [Fig.1(c)], one would expect a flatland laser to irregularly oscillate in clusters of modes without any synchronization, or in any case to be likely to show multimode oscillation. Indeed, even in the recently-introduced class of semiconductor topological insulator lasers \cite{r28}, the robustness of lasing topological states is not immune to dynamical instabilities and to the existence of attractors with complex temporal dynamics \cite{r30}.  While
such a kind of dynamical instabilities are expected to arise in semiconductor flatland lasers as well, what is rather surprisingly is that stable oscillation in a single supermode with global phase locking is not fully prevented and can be often observed after transient laser switch on. For gain provided by quantum well semiconductors \cite{r27,r28}, the dynamics is described by the class-B laser model \cite{r27,r30,r32,r33,r34}
     
\begin{figure}[htb]
\centerline{\includegraphics[width=8.6cm]{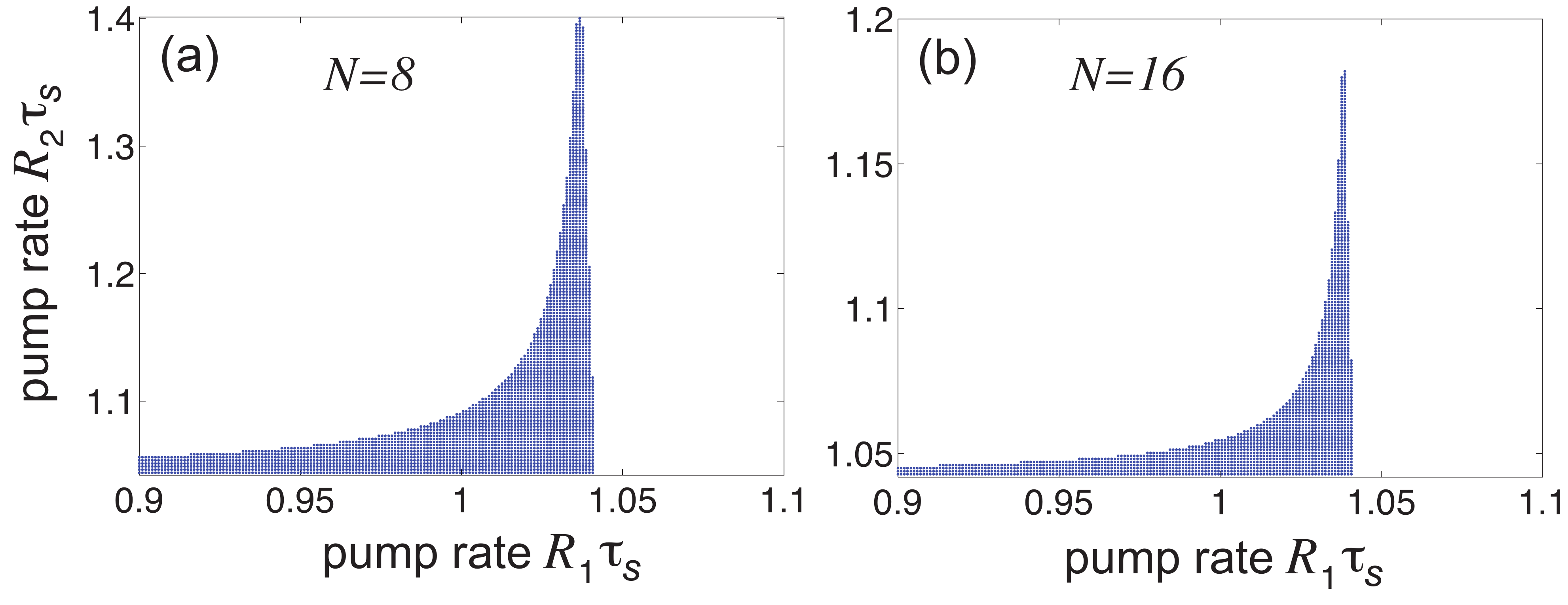}} \caption{ \small
(Color online) Stability domain (dark area) in the $(R_1 \tau_s,R_2 \tau_s)$ plane of the supermodes with Bloch wave number $q=0, \pi$ in a flabland lattice with periodic boundary conditions comprising (a) $N=8$, and (b) $N=16$ unit cells. Parameter values are: $\sigma \tau_p=24$, $\alpha=3$, $\tau_s / \tau_p=100$, and $\kappa \tau_p=2$. All other supermodes with wave number $q \neq 0, \pi$ are unstable. The threshold pump parameter is $R_{th} \tau_s=1+1/ (\sigma \tau_p) \simeq 1.0417$.}
\end{figure}  
 
  \begin{figure}[htb]
\centerline{\includegraphics[width=8.6cm]{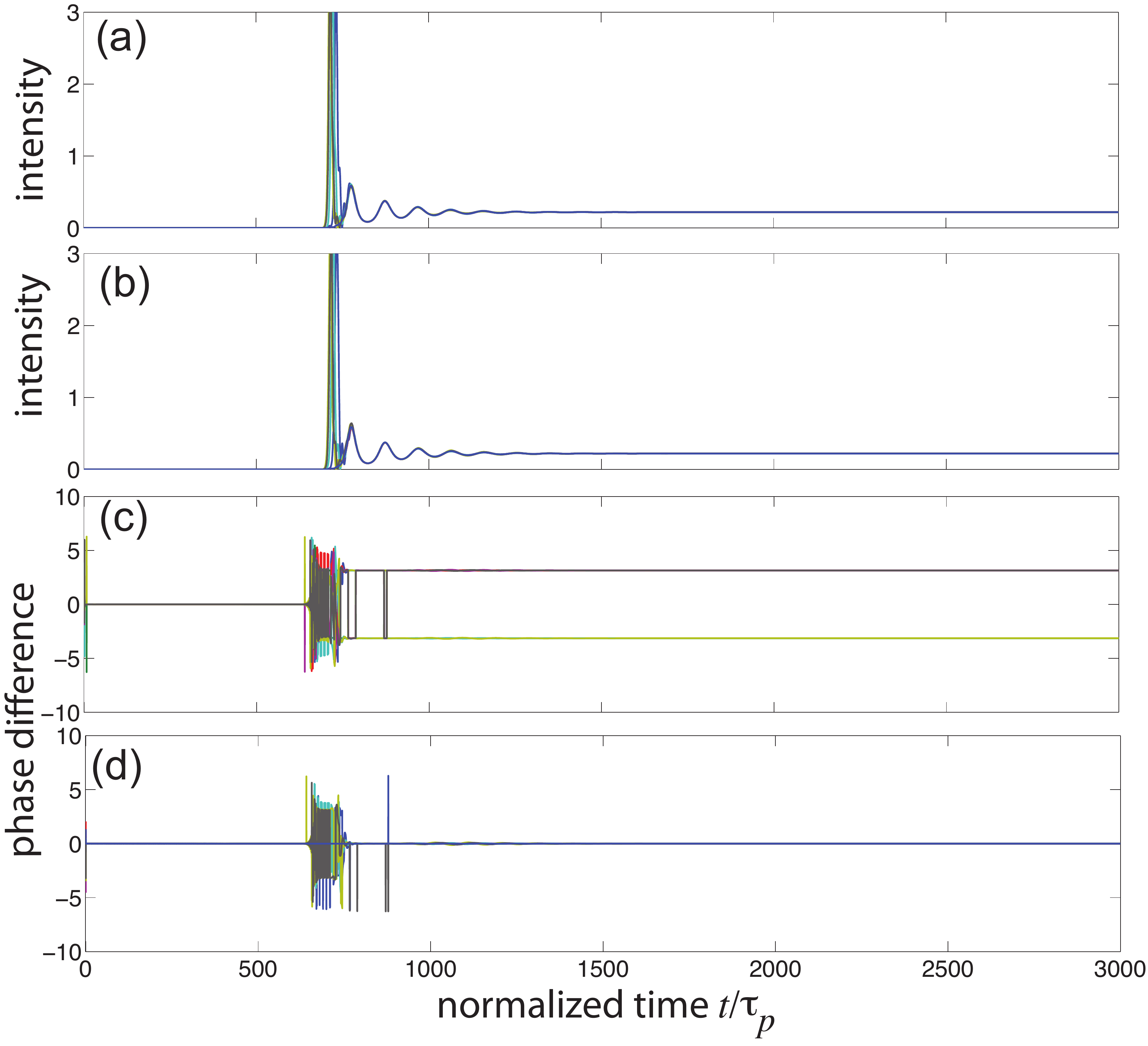}} \caption{ \small
(Color online) Switch on dynamics in a flatland laser for $N=8$, $R_1 \tau_s=1.03$ and $R_2 \tau_s=1.06$. Other parameters are as in Fig.2. Initial conditions are $N_n^A=N_n^B=N_n^C=0$ and $a_n$, $b_n$, $c_n$ random complex numbers of small mean amplitude ($ \sim 10^{-7}$). (a,b) Temporal evolution of intensities $|b_n(t)|^2$ and $|c_n(t)|^2$, respectively, for $n=1,2,...,8$. (c) Temporal evolution of the phase difference $\Delta \varphi_n=\varphi_{n+1}(t)-\varphi_n(t)$ ($n=1,2,...,7$), where $\varphi_n(t)$ is the phase of $b_n(t)$. (d) Temporal evolution of the phase difference $\theta_n(t)-\varphi_n(t)$ ($n=1,2,...,8$), where $\theta_n(t)$ is the phase of $c_n(t)$. After transient relaxation oscillations, lasing in the supermode with $q=\pi$ and $B=C$ is clearly observed. }
\end{figure}  
\begin{figure}[htb]
\centerline{\includegraphics[width=8.6cm]{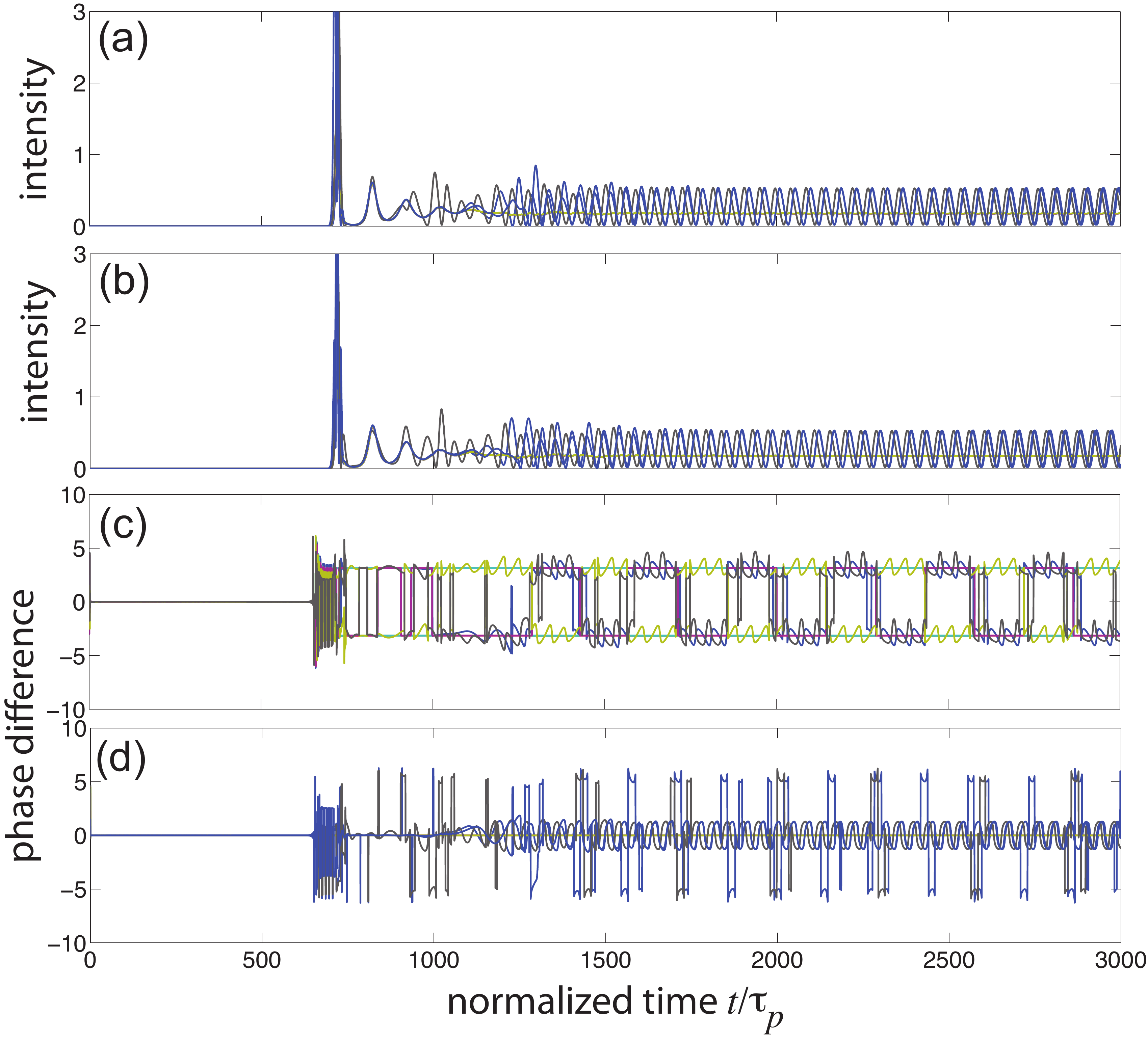}} \caption{ \small
(Color online) Same as Fig.3, but for different initial condition.}
\end{figure}  
  \begin{figure}[htb]
\centerline{\includegraphics[width=8.6cm]{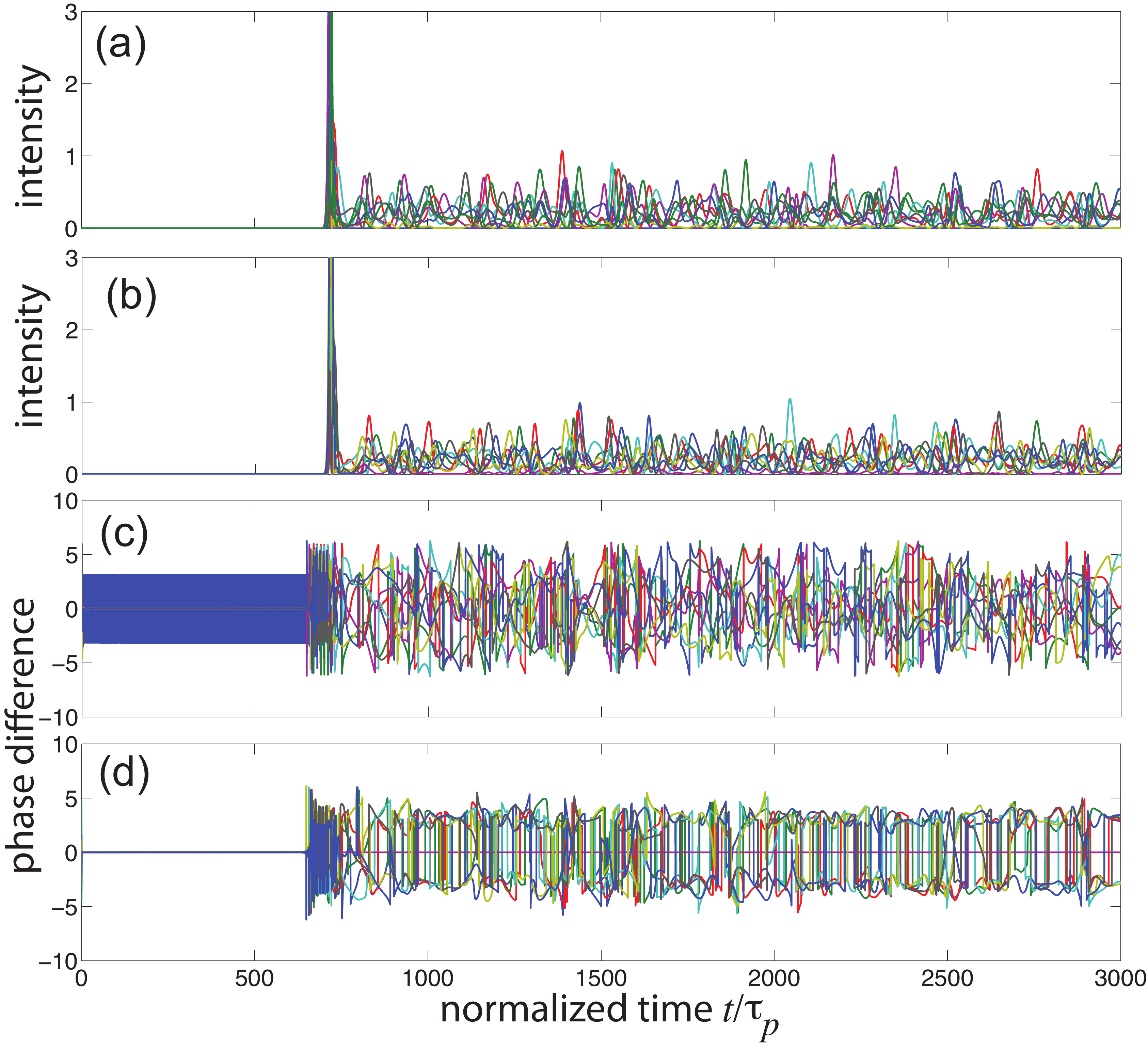}} \caption{ \small
(Color online) Same as Fig.3, but for different initial condition.}
\end{figure}  
  \begin{figure}[htb]
\centerline{\includegraphics[width=8.6cm]{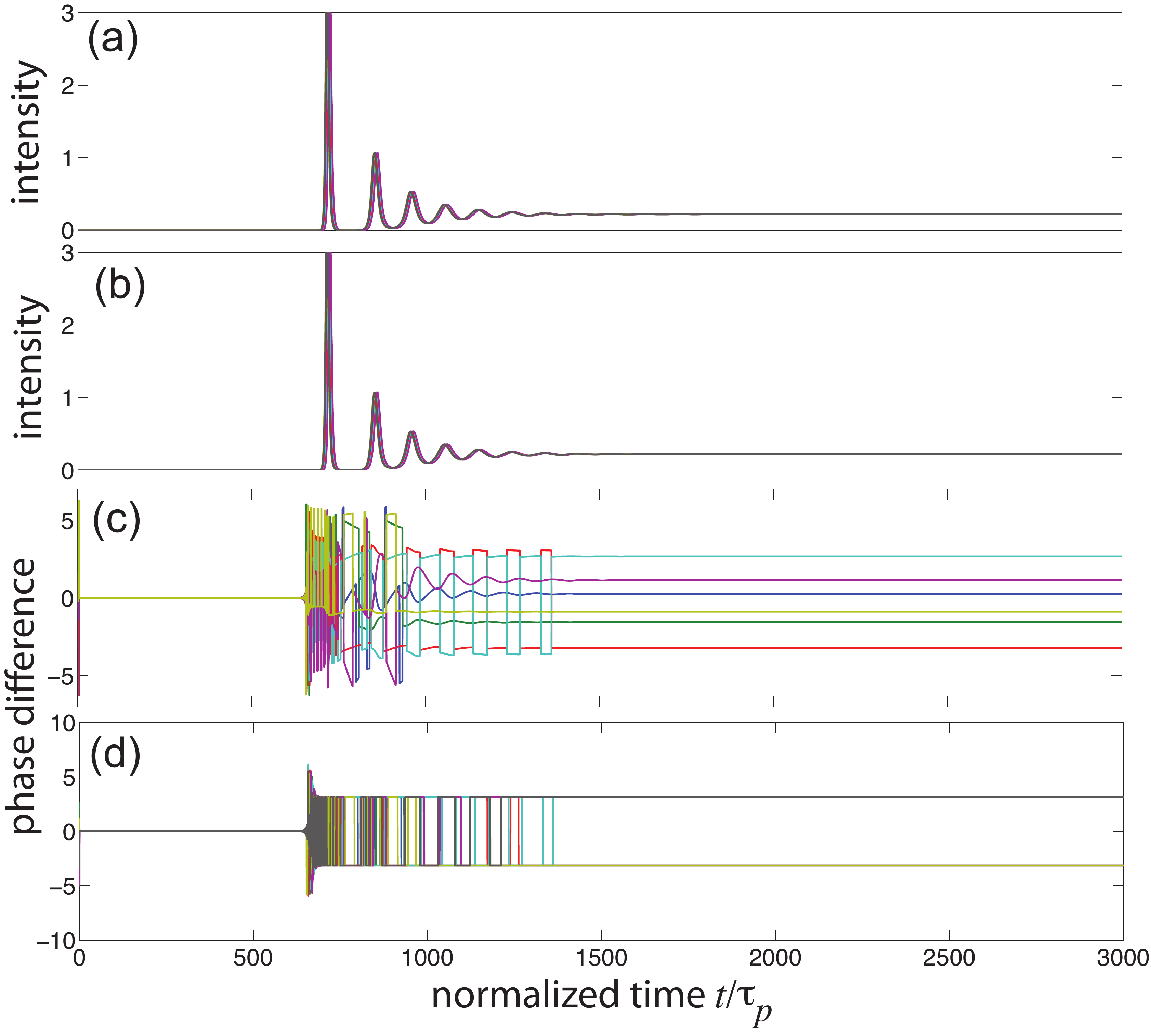}} \caption{ \small
(Color online) Same as Fig.3, but for $N=7$.}
\end{figure}  

\begin{eqnarray}
i \frac{da_n}{dt} & = & \kappa  (b_{n-1}+b_n+c_{n-1}+c_n) \nonumber \\
& + & \frac{i(1-i \alpha)}{2} \left[ -1/ \tau_p+ \sigma  (N_n^A-1) \right] a_n \nonumber \\
i \frac{db_n}{dt} & = & \kappa  (a_{n+1}+a_n) \nonumber \\
& + & \frac{i(1-i \alpha)}{2} \left[ -1/ \tau_p+ \sigma (N_n^B-1) \right] b_n \nonumber \\
i \frac{dc_n}{dt} & = & \kappa (a_{n+1}+a_n)  \\
& + & \frac{i(1-i \alpha)}{2} \left[ -1/ \tau_p+ \sigma  (N_n^C-1) \right] c_n \nonumber \\
\frac{dN^X_n}{dt} & = &  R^X -N_n^X/ \tau_s -2(N_n^X-1)|x_n|^2 / \tau_s 
 \nonumber 
\end{eqnarray}
($X=A,B,C$, $x=a,b,c$). In the above equations, $N_n^X$ is the carrier density at
the $n-th$ site of sublattice $X$, $\tau_p$ is the photon cavity lifetime, $\sigma$ is proportional
to the differential gain, $\alpha$ is the linewidth enhancement factor, $R^X$ is the normalized (sublattice-dependent) pumping rate, and $\tau_s$ is the carrier lifetime. The pump rate at threshold for a single microring is given by $R_{th}=(1/ \tau_s)(1+1/ \sigma \tau_p)$. We assume $R^A \equiv R_1<R_{th}$, $R^B=R^C \equiv R_2>R_{th}$, corresponding to microrings in sublattice A (B,C) below (above) threshold, the linear (small-signal) gain/loss parameters being given by $g=-1/ \tau_p+\sigma(R_2 \tau_s-1)$ and $\gamma=1/ \tau_p+\sigma(1-R_1 \tau_s)$. The laser equations (3) can display different types of stationary states as well as complex temporal attractors, largely dependent on laser parameters and initial conditions.
The simplest family of stationary states, corresponding to lasing states in the flatland supermodes described above, is given by
\begin{eqnarray}
a_n=0\;, \; b_n=B \exp(iq_ln), \; c_n=C \exp(iq_ln), \nonumber \\
 N^{A}_n=R_1 \tau_s \equiv N_1 \; , \; N_n^{B,C}=1+\frac{1}{\sigma \tau_p} \equiv N_2
\end{eqnarray}
where $l$ is the quantized supermode index and $|B|^2=|C|^2=[ \sigma \tau_p(R_2 \tau_s-1)-1]/2$, with the further constraint $B=-C$ for $q_l \neq \pi$.   
  The stability of the $^{\prime}$nonlinear$^{\prime}$ supermodes (4) can be performed by a standard linear stability analysis \cite{r29,r30,r33,r34}. The matrix of the linearized equations, that describe the growth of small perturbations, is a $8 \times 8$ matrix containing the Bloch wave number $Q_j=2 \pi j /N$ of the perturbation with growth rate $\lambda(Q_j)$, with $j=0,1,...,N-1$. The supermode solution (4) is linearly stable provided that the real part of $\lambda(Q_j)$ is smaller or equal to zero for any perturbation wave number $Q_j$. We numerically computed the growth rates and stability domains in parameter space $(R_1 \tau_s, R_2 \tau_s)$ for parameter values that typically apply to semiconductor quantum-well microrings \cite{r27} $\tau_s=4$ ns, $\tau_s/ \tau_p=100$, $\sigma \tau_p=24$, $\alpha=3$ and $\kappa \tau_p=2$. 
  The stability of supermodes is largely affected by the number $N$ of units cells, and a different scenario is found for $N$ even or $N$ odd.
For an even value of $N$, i.e. when the periodic boundary conditions allow for the existence of the out-of-phase supemode with $q_l= \pi$, all supermodes are unstable, except those with $l=0$ and $l=N/2$, corresponding to microrings oscillating either in phase ($q_l=0$) or with alternating phases $(q_l=\pi$). Moreover, for $q_l=\pi$ the phases $B= \pm C$ are the only stable ones. A typical behavior of the stability domain of the in-phase and out-of-phase supermodes in the $(R_1 \tau_s, R_2 \tau_s)$ plane for $N$ even is shown in Fig.2. As a general feature, the stability domain shrinks as the number of sites $N$ increases. Clearly, the stability domain is limited by two boundaries. The sharp boundary on the right side, delimited by the curve $R_1 \tau_s \simeq R_{th} \tau_s$, is simply related to the fact that microrings in sublattice A reach threshold for oscillation. On the other hand, the other stability boundary, observed as $R_2 \tau_p$ is increased at a fixed value of $R_1 \tau_s< R_{th} \tau_s $, corresponds to a Hopf-type instability driven by a non-vanishing value of the linewidth enhancement factor $\alpha$ \cite{r32,r33,r34} (this instability disappears in the $\alpha \rightarrow 0$ limit).\\
For an odd number $N$ of unit cells, i.e. when geometric frustration prevents existence of the out-of-phase supermode $q_l=\pi$, all supermodes are stable in some domain, indicating that the system is more likely to show multimode oscillation when laser is started from initial small random noise.\par
 The laser switch-on dynamics has been simulated by numerical solution of the laser rate equations (3) starting from small random complex amplitudes of modes. We assumed a flatland lattice with $N=8$ unit cells. After initial relaxation oscillation transient, different  attractors, corresponding to different dynamical regimes, can be observed depending on parameter values and initial conditions, i.e.,
different runs starting from small random noise can result in different (regular or irregular) dynamical behaviors. This is
a clear signature of highly nonlinear dynamics, which
is very common in coupled nonlinear oscillator models \cite{r34,r35}.  Over 100 runs starting from small random noise amplitudes, in 15 runs we observed stable steady-state oscillation in the $q_l=\pi$ supermode, the phase $B=C$ being the most likely one, while in other cases we observed partial phase locking and oscillations in clusters or more complex attractors. As an example, Figs.3, 4 and 5 show numerical results obtained from different small-amplitude initial conditions, leading to stable oscillation (Fig.3), partial phase-locking mixed with oscillatory dynamics (Fig.4), and highly-irregular temporal dynamics (Fig.5). 
 In Fig.3, after transient relaxation oscillations the attractor of laser dynamics is the $q_l=\pi$ supermode with $B=C$, i.e. microrings in the same unit cell oscillate in phase, whereas adjacent rings in the same sublattice (either sublattice B or C) oscillate with opposite phases. In Fig.4, steady-state oscillation and phase locking is incomplete, i.e. it is observed for a fraction of microrings in the two sublattices B and C, while oscillatory dynamics (in both amplitude and phase) is found for the other microrings. Finally, a highly irregular dynamics is observed in Fig.5. In all cases microrings A are not lasing.\\  
 For comparison, we numerically studied laser switch-on dynamics from small random noise in a flatland lattice with an odd number $N=7$ of unit cells. In this case, steady-state emission is rarely observed (5 times over 100 runs). Even when steady-state oscillation in all microrings of sublattices A and B is observed, phase locking arises in clusters of A/B dimers, i.e. the phases of modal fields between adjacent unit cells $(n,n \pm 1)$ are not locked. An example of steady-state oscillation, without global phase locking, is shown in Fig.6.   Our results thus show that, provided that the number of unit cells in the lattice is even, the flatland lattice can coherently emit steady-state radiation with phase locking among all the microrings, albeit attractors corresponding to complex dynamical behaviors could be observed like in topological insulator lasers \cite{r30}.\par
 {\it Conclusions.} Topological and flatland photonic structures are currently hot areas of research in optics. While most studies focused on passive structures, recent works suggested that the introduction of selective gain and loss can be of potential relevance, for example for the realization of topological insulator lasers \cite{r28}, where laser emission occurs in a topologically-protected edge state. Here we introduced the idea of flatland laser, where laser emission  occurs in a flat band of a lattice. In spite of the high-degree of threshold degeneracy, we found that stationary laser emission in a single supermode with global phase locking can be observed. However, like in topological insulator lasers \cite{r30}, we observed dynamical instabilities and complex temporal attractors. These could be mitigated by non-Hermitian coupling engineering \cite{r34}, if needed. Cooperative lasing emission in flat band structures provides an interesting route toward the realization of high-power stable laser emission in lattice systems compatible with current integrated semiconductor laser technology. There are some open questions, that could motivate further investigations beyond laser-oriented applications. For example, how does a flat band in a network system influence the formation of complex states such as chimera or rogue-wave states? In case of photonic lattices sustaining more than one flat band \cite{r36}, is it possible to achieve global synchronization in the system?

\newpage

%%%%%%%%%%%%%%%%%%%%%%%%%%%%%%%
% References with full titles %
%%%%%%%%%%%%%%%%%%%%%%%%%%%%%%%


\begin{thebibliography}{99}


%%%%%%%%%%%%%%%%%%%%%%%%%%%%%%%
% References (short version)  %
%%%%%%%%%%%%%%%%%%%%%%%%%%%%%%%

\bibitem{r1}
D. Leykam and S. Flach, APL Photon. {\bf 3}, 070901 (2018).
\bibitem{r2}
D. Leykam, A. Andreanov, and S. Flach, Adv. Phys. X {\bf 3}, 1473052 (2018).
\bibitem{r3}
D. Leykam, S. Flach, O. Bahat-Treidel, and A.S. Desyatnikov, Phys. Rev. B {\bf 88},  224203 (2013).
%S. Flach, D. Leykam, J.D. Bodyfelt, P. Matthies, and A.S. Desyatnikov, EPL {\bf 105}, 30001 (2014).
\bibitem{r4}
A. Crespi, G. Corrielli, G. Della Valle, R. Osellame, and S. Longhi, New J. Phys. {\bf 15}, 013012 (2013).
\bibitem{r5}
D. Guzm\'an-Silva, C. Mejia-Cort\'es, M.A. Bandres, M.C. Rechtsman, S. Weimann, S. Nolte, M. Segev, A. Szameit, and R.A. Vicencio, New J. Phys. {\bf 16}, 063061 (2014).
\bibitem{r6}
S. Longhi, Opt. Lett. {\bf 39}, 5892-5895 (2014).
\bibitem{r7}
R.A. Vicencio, C. Cantillano, L. Morales-Inostroza, B. Real, C. Mej\'{\i}a-Cort\'es, S. Weimann, A. Szameit, and M.I. Molina, Phys. Rev. Lett. {\bf 114}, 245503  (2015).
\bibitem{r8}
S. Mukherjee, A. Spracklen, D. Choudhury, N. Goldman, P. \"{O}hberg, E. Andersson, and R.R. Thomson, Phys. Rev. Lett. {\bf 114}, 245504 (2015).
\bibitem{r9}
S. Mukherjee and R.R. Thomson, Opt. Lett. {\bf 40}, 5443-5546 (2015).
\bibitem{r10}
 Y. Zong, S. Xia, L. Tang, D. Song, Y. Hu, Y. Pei, J. Su, Y. Li, and Z. Chen, Opt. Express {\bf 24}, 8877-8885 (2016).
 \bibitem{r11}
S. Xia, Y. Hu, D. Song, Y. Zong, L. Tang, and Z. Chen, Opt. Lett. {\bf 41}, 1435-1438 (2016).
\bibitem{r12}
S. Mukherjee, A. Spracklen, M. Valiente, E. Andersson, P. \"{O}hberg, N. Goldman, and R.R. Thomson, Nat. Commun. {\bf 8},  13918 (2017).
\bibitem{r13}
B. Real, C. Cantillano, D. L\'opez-Gonz\'alez, A. Szameit, M. Aono, M. Naruse, S.-J. Kim, K. Wang, and R.A. Vicencio,  Sci. Rep. {\bf 7}, 15085 (2017).
\bibitem{r14}
S. Mukherjee, M. Di Liberto, P. \"{O}hberg, R.R. Thomson, and N. Goldman, Phys. Rev. Lett. {\bf 121}, 075502 (2018).
\bibitem{r15}
M. Kremer, I. Petrides, E. Meyer, M. Heinrich, O. Zilberberg, and A. Szameit, arXiv:1805.05209 (2018).
\bibitem{r16}
B. Sutherland, Phys. Rev. B {\bf 34}, 5208-5211 (1986).
\bibitem{r17}
E. H. Lieb, Phys. Rev. Lett. {\bf 62}, 1201-1204 (1989).
\bibitem{r18}
L. Ge, Phys. Rev. A {\bf 92}, 052103 (2015).
\bibitem{r19}
D. Leykam, S. Flach, and Y. D. Chong, Phys. Rev. B {\bf 96}, 064305 (2017).
\bibitem{r20}
B. Qi, L. Zhang, and L. Ge, Phys. Rev. Lett. {\bf 120}, 093901 (2018).
\bibitem{r21}
L. Ge, Photonics Res. {\bf 6}, A10 (2018).
\bibitem{r22}
L. Pilozzi and C. Conti, Phys. Rev. B {\bf 93}, 195317 (2016).
\bibitem{r23}
L. Pilozzi and C. Conti, Opt. Lett. {\bf 42}, 5174 (2017).
\bibitem{r24}
B. Bahari, A. Ndao, F. Vallini, A. El Amili, Y. Fainman, and B. Kante, Science {\bf 358},  636 (2017).
\bibitem{r25}
P. St-Jean, V. Goblot, E. Galopin, A. Lemaitre, T. Ozawa, L. Le Gratiet, I.Sagnes, J. Bloch, and A. Amo, Nat. Photon. {\bf 11}, 651-656 (2017).
\bibitem{r26}
H. Zhao, P. Miao, M.H. Teimourpour, S. Malzard, R. El-Ganainy, H. Schomerus, and L. Feng, Nat. Commun. {\bf 9}, 981 (2018).
\bibitem{r27}
M. Parto, S. Wittek, H. Hodaei, G. Harari,  M.A. Bandres, J. Ren, M.C. Rechtsman, M. Segev, D.N. Christodoulides, and M. Khajavikhan, Phys. Rev. Lett. {\bf 120}, 113901 (2018).
\bibitem{r28}
M.A. Bandres, S. Wittek, G. Harari, M. Parto, J. Ren, M. Segev, D.N. Christodoulides, and M. Khajavikhan, Science {\bf 359}, eaar4005 (2018).
\bibitem{r29}
S. Longhi, Ann. Phys. (Berlin) {\bf 530}, 1800023 (2018).
\bibitem{r30}
S. Longhi, Y. Kominis and V. Kovanis, EPL {\bf  122}, 14004 (2018).
\bibitem{r30bis}
S. Malzard and H. Schomerus, New J. Phys. {\bf 20}, 063044 (2018).
\bibitem{r30tris}
S. Malzard, E. Cancellieri, and H. Schomerus, Opt. Express {\bf 26},  22506-22518 (2018).
\bibitem{r31}
D.A. Smirnova, P. Padmanabhan, and D. Leykam (unpublished).
\bibitem{r32}
Y. Kominis, V. Kovanis, and T. Bountis, Phys. Rev. A {\bf 96}, 043836 (2017).
\bibitem{r33}
R. D. Li and T. Erneux, Phys. Rev. A {\bf 46}, 4252 (1992).
\bibitem{r34}
S. Longhi and L. Feng,  APL Photon. {\bf 3}, 060802 (2018).
\bibitem{r35}
K. Wiesenfeld and P. Hadley, Phys. Rev. Lett. {\bf 62}, 1335 (1989).
\bibitem{r36}
C. Cantillano, S. Mukherjee, L. Morales-Inostroza, B. Real, G. C\`aceres-Aravena, C. Hermann-Avigliano, R.R. Thomson, and R.A. Vicencio, New J. Phys. {\bf 20}, 033028 (2018). 

 \end{thebibliography}
\end{document}